# Regarding the Accretion of 2003 $VB_{12}$ (Sedna) and Like Bodies in Distant Heliocentric Orbits

S. Alan Stern
Department of Space Studies
Southwest Research Institute
1050 Walnut Street, Suite 400
Boulder, CO 80302 USA

astern@swri.edu
[303]546-9670 (voice)
[303]546-9687 (fax)

12 October 2004

## Abstract

Recently, Brown et al. (2004) reported the exciting discovery of an ~800 km radius object, (90377) Sedna, on a distant, eccentric orbit centered at ~490 AU from the Sun. Here we undertake a first look exploring the feasibility of accreting this object and its possible cohorts between 75 AU (Sedna's perihelion distance) and 500 AU (Sedna's semi-major axis distance) from the Sun. We find such accretion possible in a small fraction of the age of the solar system, if such objects were initially on nearly circular orbits in this region, and if the solar nebula extended outward to distances far beyond the Kuiper Belt. If Sedna did form in situ, it is likely to be accompanied by a cohort of other large bodies in this region of the solar system.

# 1. Introduction

The discovery of the large (R~800 km), distant (q~75 AU, a~490 AU) object, (90377) Sedna = 2003 VB12 (Brown, Trujillo, and Rabinowitz 2004), raises numerous interesting questions regarding the origin and evolution of our solar system. Among these is the question of where Sedna could have accreted.

Based on discovery statistics, Sedna's discoverers (Brown et al. 2004; Bea04) suggest that Sedna is in the inner reaches of the Oort Cloud, meaning it was scattered outward from a site much closer to the Sun, for example (i.e., the giant planets region) to its present orbit either by planetary or other (e.g., stellar) perturbations. Morbidelli & Levison (2004; ML04) make a similar, implicit assumption about Sedna's origin having been in the region of the giant planets and having been scattered outward to its present semi-major axis.

Clearly, Sedna cannot have formed it its present, eccentric and inclined orbit, since the ~1 km $sec^{-1}$ random encounter velocities characteristic of such an orbit are too high to allow accretion. This easy to reach finding makes clear that Sedna's orbit must be evolved. However, this does not equate to the conclusion that Sedna must have formed in the giant planets region, and then been scattered outward, or whether instead Sedna might have formed at large heliocentric distances, beyond the orbits of the giant planets, having then been perturbed to its present eccentricity and inclination in a manner such as those described by Bea04 and ML04. Here we examine the feasibility of accretion at large heliocentric distances characteristic of Sedna's present orbit, in order to determine whether the excitation of Sedna's e and I must have been accompanied by a large increase in its a.

# 2. Accretion Model

In order to investigate the accretion of Sedna-like objects, we used a well-established, single-zone, time-dependent accretion-collisional evolution model first developed some years ago (Stern & Colwell 1997a,b). This model is based on a "moving-bin" mass evolution formalism originally developed by Wetherill (1990). The mass in the heleocentric zone of interest is tracked in a suite of 81 logarithmic mass bins separated by factors of 2 in mass. We follow objects as small as 0.01 km in radius to as large as 2000 km in radius.

At each time step the model computes the number of collisions and the resulting outcomes of these collisions occurring in an annular zone at the specified heliocentric distance, which is embedded in a disk-like solar nebula. Both the amount of debris generated by energetic collisions, as well as the degree of growth resulting from gentler collisions, are tracked as a function of target body mass, and fed back to evolve the differential population-mass structure at each time step. During runs, the code dynamically selects a time step that limits the rate of change of the fastest-changing mass bin to <1%. The model typically conserves mass to a few parts in $10^{-15}$ or better after a run time of 4.5 Gyr.

The collision frequency calculations are accomplished using a locally averaged, statistical particle-in-a-box formalism (e.g., Lissauer & Stewart 1993). Collision cross sections were appropriately scaled for gravitational focusing, but limited by both Keplerian shear and three-body effects (e.g., Ward 1996). For each impact, the model initially adds the mass of the impactor to the target and then removes the amount of debris, which completely escapes (if any, depending on the collision circumstances).

Depending on the energetics of the collisions and the masses and mechanical properties of the colliding bodies, each collision results in either some amount of net erosion (up to catastrophic fragmentation) or some amount of net accretion to the collision target (up to complete merger of the impactor). The result is net accretion to the target body if the mass of the escaping ejecta is less than the impactor mass. Objects smaller than the smallest discrete model bin tracked (10 meters in the runs presented below) are placed into a non-interacting bin to track the total mass of "debris" particles.

The code computes the amount of debris, which escapes the gravitational potential of each collision pair, using established scaling relations applied in a wide range of circumstances (Housen & Holsapple 1990; hereafter HH90). In the runs reported on below, a material strength of $s=3x10^5$ erg $cm^{-3}$ was assumed; this strength parameter is applicable to ice. Other strength values (e.g., both weaker values representative of sand and snow and stronger values representative of competent rock) will be evaluated in a future paper exploring a more extensive parameter space of accretion runs at large heliocentric distances.

The amount of debris resulting from any given collision is estimated as follows. For each collision pair of mass $m_K<m_L$ colliding at relative velocity v, the specific impact energy is computed according to the standard definition, $Q=(½m_Kv^2)/m_L$. Q is then compared to a threshold value for catastrophic shattering, called $Q_S^*$. The code uses a strain-rate scaling model for $Q_S^*$ (HH90); check runs made during code validation using an energy-scaling model (Davis et al. 1985) resulted in only minor differences in run results.

If $Q>Q_S^*$, then the mass fraction with escape velocity from the colliding pair is given by $f(>v_{esc})=½(v_{esc}/v_{med})^{3/2}$, where $v_{med}=(2f_{KE}Q)^{1/2}$ is the median fragment velocity, $f_{KE}$ is the fraction of impact energy partitioned into fragment kinetic energy, and $v_{esc}$ is the escape velocity. Experimental results indicate $f_{KE}$ is plausibly between 0.05 and perhaps 0.15 (e.g., Fujiwara et al. 1989; Arakawa et al. 1996; Asphaug, priv. comm.); we took $f_{KE}$=0.10 for the runs described below.

Because the fragmentation behavior of the bodies involved in mutual collisions is uncertain, we explored three different fragmentation properties: un-bonded sand, bonded sand, and competent (basalt) rock. The debris resulting from any given collision can result in any outcome from complete accretion (no debris achieves escape velocity from the colliding pair), to complete erosion (more than half the mass of the original target has escape velocity from the colliding pair). The escaping debris is partitioned to the mass bins below the size of the target following a standard, two-component power-law size distributions computed from various material properties derived from laboratory experiments.

In "cratering impacts" (i.e., $Q<Q_S^*$), the total ejecta mass is computed from the impactor energy following standard techniques (e.g., Colwell & Esposito 1990). The debris size distribution is then computed as a single-valued, cumulative power law ($n(>m) \sim m^{-5/6}$), with a fragment velocity distribution cumulative power-law exponent in mass of –1.2 (for unconsolidated target material runs) and –2.0 (for consolidated target material runs), consistent with laboratory experiments on hypervelocity impacts into sand and basalt.

The mean random velocity of the bodies in the run zone evolves with time in a manner consistent with the evolution of the mass spectrum, so that the largest bodies stir up the smaller bodies, and the smaller bodies damp the mean random velocities of the larger bodies owing to dynamical friction (e.g., Wetherill & Stewart 1989, 1993). A computationally fast velocity evolution scheme that ensures energy equipartition across the size bins, a key result of Wetherill and Stewart's work is used; the code assumes <i>=½<e> as a computational convenience.

## 3. Accretion Simulations at 75 and 500 AU

We used the accretion model described above (see also Stern & Colwell 1997a,b) to conduct a suite of initial accretion simulations at 75 AU and 500 AU—Sedna's perihelion and semi-major axis distances. These distances span much of the likely range of potential origin sites for Sedna outside the orbits of the giant planets.

In the runs undertaken for this first study, the heliocentric radial zone width over which we average input parameters like orbital eccentricity and spatial number density was typically ±15% of the zone's central heliocentric distance.

A surface mass density of 0.1 g $cm^{-2}$ at 40 AU corresponds to a minimum mass solar nebula. Surface mass densities $\Sigma$=0.05, 0.12, and 0.20 g $cm^{-2}$ were assumed at 40 AU. These surface mass densities were propagated out to our run distances of 75 AU or 500 AU assuming a heliocentric disk that varies like $R^{-2}$. Higher surface mass densities at 40 AU and shallower falloffs with heliocentric distance, as argued for by some (e.g. Lissauer 1987), increase $\Sigma$ at any given location, thereby promoting shorter growth timescales of large objects from planetesimals. We chose not to use these in order not to bias the results toward faster accretion.

The three adopted 40 AU surface mass densities respectively correspond to primordial 30-50 AU zone Kuiper Belt masses of approximately 09, 23, and 38 $M_{Earth}$; this is consistent with the 10 to 50 $M_{Earth}$ called for by accretion code studies of the Kuiper Belt (Stern & Colwell 1997; Davis & Farinella 1997; Kenyon 2002; see also the reviews by Farinella et al. 2000 and Stern & Kenyon 2003).

Each run started with an initial population consisting exclusively of planetesimal building blocks 1 to 10 kilometers in radius. All bodies were assumed to have a density of 1.5 g $cm^{-3}$ for the runs at 75 AU. Initial random velocities were set to a value of $10^{-3}V_K$ for the building blocks, where $V_K$ is the local Kepler speed.

The suite of simulation runs we performed at 75 AU is described in Table 1; also included in Table 1 is the elapsed time at which the first 1000 km radius body was achieved. Figure 1 depicts the largest body in each simulation as a function of elapsed time. Figure 2 depicts the starting and ending size distributions for each run.

From Table 1 and Figures 1 and 2, we draw the following conclusions:

- Starting from 1-10 km seeds some 75 AU from the Sun, it is possible to grow 1000-km radius (~2x the mass of Sedna) objects under a variety of differing conditions consistent with minimum (and greater) mass solar nebulas that extend out to 80 AU or further.
- These 1000 km radius bodies grow on comparatively short timescale—2% to 20% the 4.5 Gyr age of the solar system.
- Still larger objects can be grown over longer timescales
- As expected from studies previously performed in the 30-50 AU zone, material properties play a greater role in determining the growth timescale with progressively lower surface mass densities.

It is also worth noting that within typically a few percent of the time at which the first 1000 km body was grown in these simulations, additional bodies in the zone reached this size, indicating that if Sedna formed in situ, then it is likely to be only one of many objects of this size in the region where it orbits.

In addition to the suite of runs made at 75 AU, a more limited suite of runs was made at 500 AU. These runs were meant only to test the feasibility of constructing Sedna-scale bodies at 500 AU.

The suite of simulation runs performed at 500 AU is described in Table 2. Again three surface mass densities were run. These were: $\Sigma$=0.0003 g cm$^{-2}$, 0.0022 g cm$^{-2}$, and 0.0057 g cm$^{-2}$; corresponding to 30-50 AU zone masses of 3 $M_{Earth}$, 18 $M_{Earth}$, and 47 $M_{Earth}$, propagated outward to 500 AU like $R^{-2}$. From these runs we found that:

- 1000-km radius (Sedna-class) objects could be grown at 500 AU, but only in the latter two of the three cases we ran.
- Not surprisingly, this indicates the need for a massive solar nebula extending to such distances if in situ growth is to occur there.
- As shown in Table 2, starting from 1-10 km seeds, the growth times for 1000 km radius bodies in simulations with sufficiently extended, massive disks, was on a timescale that is short compared to the age of the solar system.

## 4. Discussion

Observations of nearby debris disks revealed long ago (see Backman & Paresce 1993; see also Kenyon et al. 1999; Gladman et al. 2001) that these systems often extend outward 100s and sometimes even >1000 AU.

Here we presented a first look at the viability of accreting objects like the newly discovered, distant, ~800 km radius body 90377 Sedna. It was found that Sedna and even larger bodies could indeed have grown from km-class planetesimal seeds in a distant extension of the solar nebula. Such a distant extension of our own solar system has been speculated about in the past (e.g., Stern 1996; Farinella et al. 2000; Kenyon 2002). It was found that Sedna can indeed have accreted at great distance from the Sun, even comparable to its present orbital semi-major axis.

As we noted at the outset, although Sedna-like bodies *could* have grown at large distances characteristic of Sedna's present orbit, they cannot have accreted on its present high eccentricity, comparatively high inclination orbit, because such an orbit produces collisions that are too energetic, thereby promoting net mass erosion, rather than accretion. Hence, it is clear that (i) Sedna's dynamical history involves one or more significant evolutionary events in the past, and (ii) that such events must have postdated Sedna's formation epoch. Morbidelli & Levison (2004) argue strongly that a close stellar encounter with the solar system is the most likely culprit behind Sedna's orbital evolution.

As shown by the results presented here, it is not clear whether the event that generated Sedna's orbital eccentricity-inclination excitation also brought it from a location much closer to the Sun than it presently reaches at perihelion, or whether instead Sedna formed somewhere within its present heliocentric distance range but later had its orbit excited to the present e and I.

This is a compelling issue because a formation location for Sedna at or beyond its present perihelion location would strongly indicate that the Sun's planetesimal disk far extended well beyond the Kuiper belt. It would also indicate that the observed Kuiper Belt "edge" near 50 AU is not a final terminus, but simply the inner edge of an annual trough or gap.

To better constrain Sedna's provenance, it will be necessary to determine if there is a large cohort objects in similarly distant orbits. A particularly telling clue will be revealed when it is determined whether or not there are large bodies on orbits with q>75 AU on *nearly circular* orbits that have not been excited, and therefore were more likely to be primordial.

## Acknowledgements

Hal Levison and Alessandro Morbidelli provided interesting discussions as this work was undertaken. Luke Dones, Martin Duncan, and Dan Durda provided helpful feedback on the manuscript; an anonymous referee provided helpful comments.

**Table 1: Time to Grow a 1000 km Body at 75 AU**

| Σ (75 AU) | Fragmentation Property Assumption | Run Time |
|---|---|---|
| 0.017 g $cm^{-2}$ | UQS | $4.1x10^8$ yrs |
| 0.017g $cm^{-2}$ | BQS | $8.1x10^8$ yrs |
| 0.017 g $cm^{-2}$ | BAS | $6.1x10^8$ yrs |
| 0.034 g $cm^{-2}$ | UQS | $2.0x10^8$ yrs |
| 0.034 g $cm^{-2}$ | BQS | $2.3x10^8$ yrs |
| 0.034 g $cm^{-2}$ | BAS | $2.1x10^8$ yrs |
| 0.056 g $cm^{-2}$ | UQS | $1.1x10^8$ yrs |
| 0.056 g $cm^{-2}$ | BQS | $1.4x10^8$ yrs |
| 0.056 g $cm^{-2}$ | BAS | $1.4x10^8$ yrs |

UQS= Unbonded Quartz Sand; BQS=Bonded Quartz Sand, BAS=Basalt; see text.

**Table 2: Time to Grown a 1000 km Body at 500 AU**

| Σ (500 AU) | Fragmentation Property Assumption | Run Time |
|---|---|---|
| 0.0022 g $cm^{-2}$ | UBS | $5.7x10^8$ yrs |
| 0.0057 g $cm^{-2}$ | UBS | $1.4x10^8$ yrs |

## Figure Captions

Figure 1. The largest body in each of nine accretion simulations in the 70-80 AU zone is shown here as a function of time; the nine runs are described in §3 and Table 1. From the slope of these curves, one can easily pick out the three major regimes of accretion in these simulations—i.e., slow, binary accretion prior to gravitational runaway, steep (fast growing) gravitational runaway, and finally, slower oligarchic growth. Note that, owing to the statistical nature of the code employed, the accretion time estimates shown here can only be considered representative; in cases when multiple runs were made with the same properties, accretion times were found to vary by factors of about two.

Figure 2. The starting and ending differential population size distributions in each of nine accretion simulations in the 70-80 AU zone is shown here as a function of time; the nine runs are described in §3 and Table 1. Note that the ending size distribution is not that expected at the ultimate end of accretion, but simply a snapshot taken near the time the first 100 km bodies were formed.

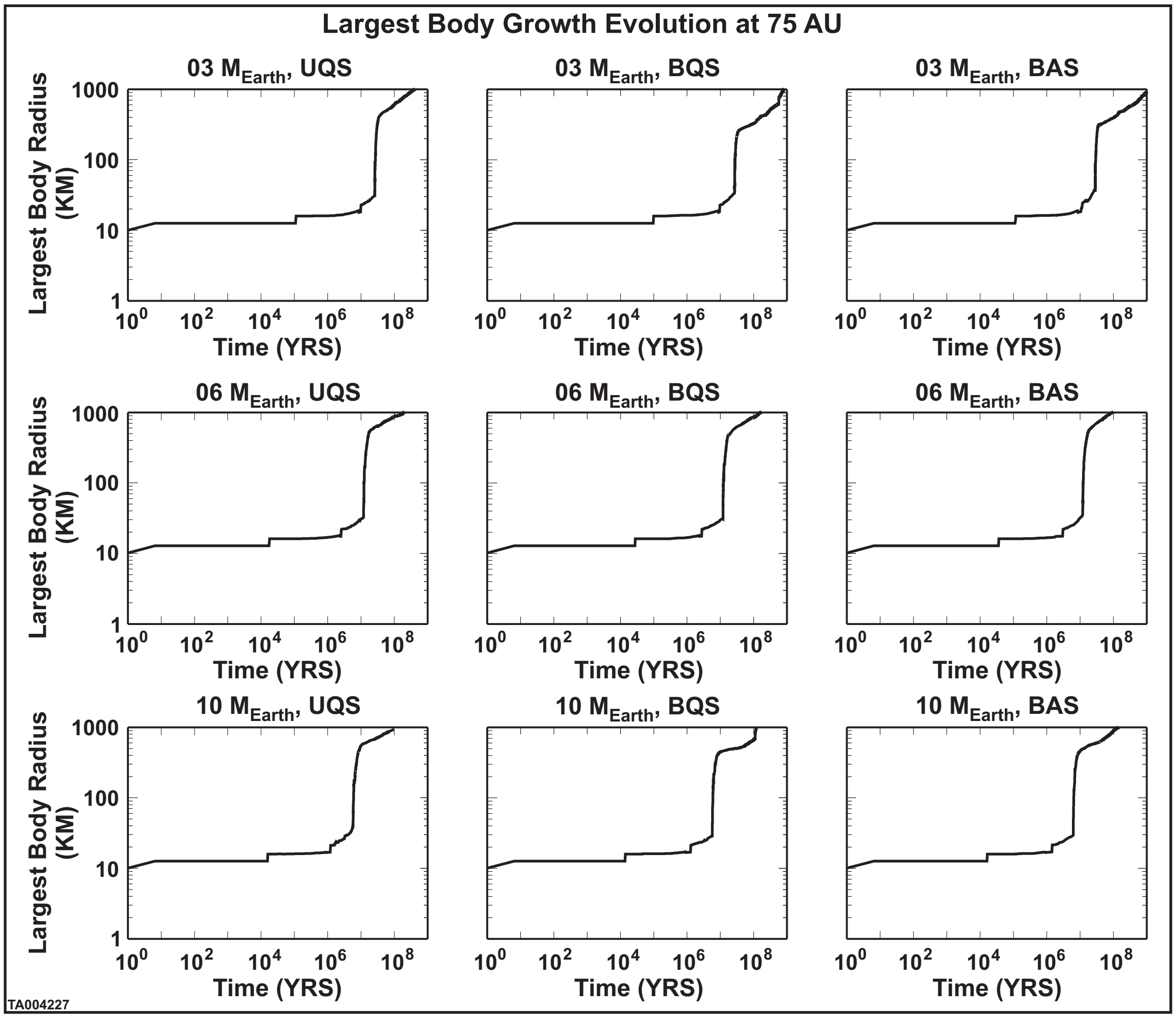

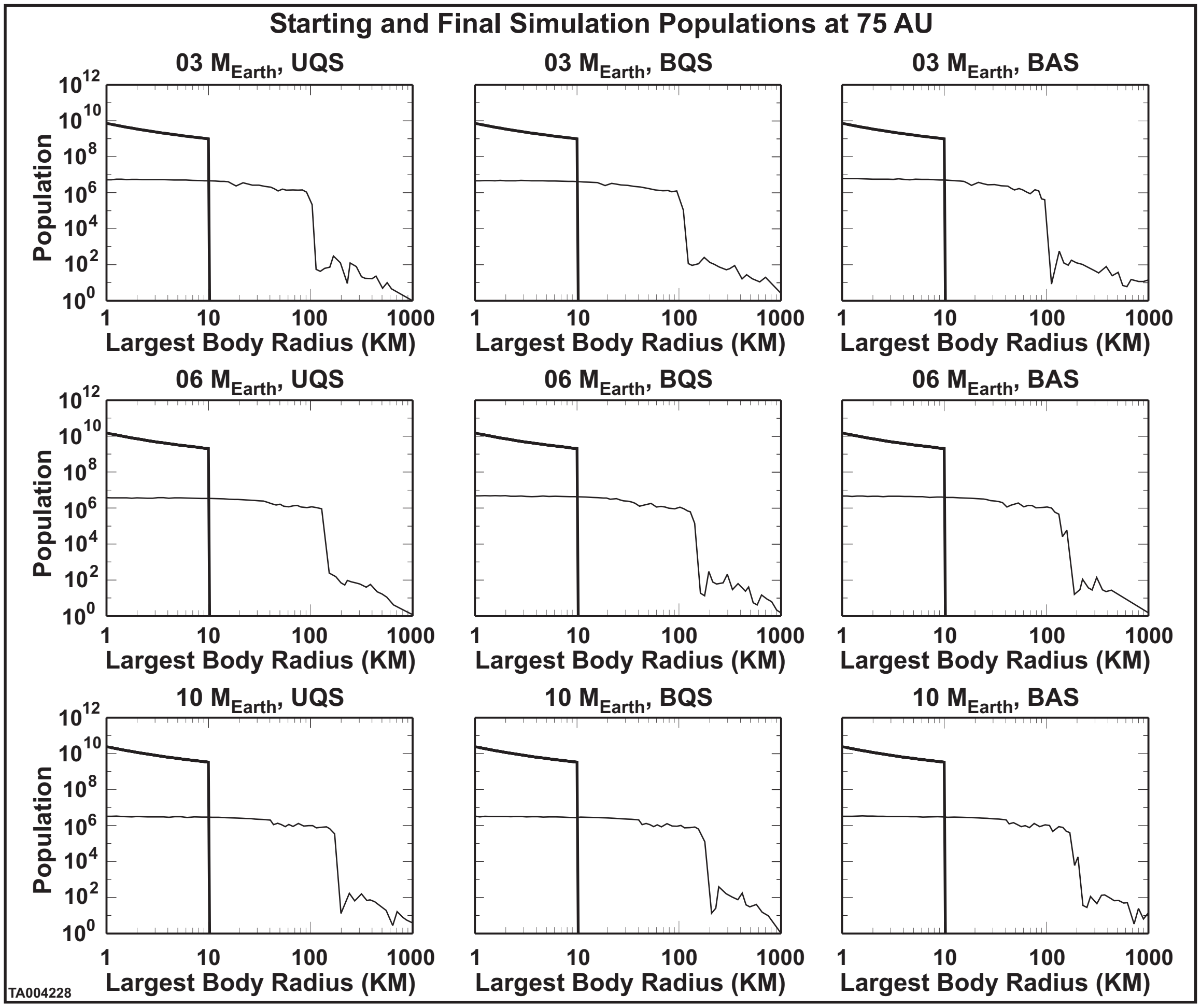

Starting and Final Simulation Populations at 75 AU
03 M_Earth, UQS
03 M_Earth, BQS
03 M_Earth, BAS
06 M_Earth, UQS
06 M_Earth, BQS
06 M_Earth, BAS
10 M_Earth, UQS
10 M_Earth, BQS
10 M_Earth, BAS
Population
10^0
10^2
10^4
10^6
10^8
10^10
10^12
Largest Body Radius (KM)
1
10
100
1000
TA004228